# Boosting classical and quantum nonlinear processes in ultrathin van der Waals materials

Xiaodan Lyu[1,2,†], Leevi Kallioniemi[1,†], Hongbing Cai[3], Liheng An[1], Ruihuan Duan[4], Shuin Jian Wu[5], Qinghai Tan[6], Chusheng Zhang[1], Ruihua He[7], Yansong Miao[7], Zheng Liu[4], Alexander Ling[5,8], Jesus Zúñiga-Perez[1,2,*], Weibo Gao[1,2,9,10,11,*]

[1]Division of Physics and Applied Physics, School of Physical and Mathematical Sciences, Nanyang Technological University; Singapore, 637371, Singapore

[2]Majulab, International Research Laboratory IRL 3654, CNRS, Université Côte d'Azur, Sorbonne Université, National University of Singapore, Nanyang Technological University, Singapore, Singapore

[3]Department of Physics, University of Science and Technology of China, Hefei, China

[4]School of Materials Science and Engineering, Nanyang Technological University, Singapore, 639798, Singapore

[5]Centre for Quantum Technologies, National University of Singapore, Singapore, Singapore

[6]School of Microelectronics, University of Science and Technology of China, Hefei, China

[7]School of Biological Sciences, Nanyang Technological University, Singapore, Singapore

[8]Department of Physics, National University of Singapore, Blk S12, 2 Science Drive 3, 117551, Singapore

[9]School of Electrical and Electronic Engineering, Nanyang Technological University, Singapore, Singapore

[10]Quantum Science and Engineering Centre (QSec), Nanyang Technological University, Singapore, Singapore

[11]Centre for Quantum Technologies, Nanyang Technological University, Singapore, Singapore

† These authors contributed equally to this work.

*Corresponding author. Email: jesus.zuniga@ntu.edu.sg; wbgao@ntu.edu.sg

Understanding and controlling nonlinear processes is crucial for engineering light-matter interaction and generating non-classical light. A significant challenge in ultra-thin nonlinear materials is the marked diminution of the nonlinear conversion efficiency due to the reduced light-matter interaction length and, in many cases, the centrosymmetric crystalline structures. Here we relax these limitations and report a giant boost of classical and quantum nonlinear processes in ultrathin van der Waals materials. Specifically, with a metal-nonlinear material heterostructure we enhance classical second-harmonic generation in h-BN flakes by two-orders of magnitude. Moreover, we have engineered a metal-$SiO_2$-nonlinear material heterostructure resulting in a remarkable two orders of magnitude augmentation of the quantum spontaneous parametric down-conversion (SPDC) in $NbOCl_2$ flakes. Notably, we demonstrate SPDC in a 16 nm-thick $NbOCl_2$ flake integrated into the proposed structure. These findings simplify on-chip quantum state engineering and accelerate the use of van der Waals materials in nonlinear optoelectronics.

# Introduction

The integration of active photonic materials such as III-V semiconductors onto complementary metal-oxide semiconductor (CMOS) and dielectric photonic circuits has faced longstanding challenges. In this context, the ease of exfoliation and transfer of van der Waals materials onto nearly any material platform has driven the development of integrated photonic devices based thereon. The use of van der Waals materials as integrated nonlinear devices began more than ten years ago, when graphene was exploited for fabricating modulators and photodetectors [1-3]. More recently, the advent of transition metal dichalcogenides (TMDs) has further enriched this field. Nonlinear processes have been demonstrated in TMDs [4-9], thus making it possible to realize ultrafast optical switching [10, 11]. Beyond classical nonlinear processes, van der Waals materials have been reported to facilitate optical parametric amplification and oscillation [12, 13], and spontaneous parametric down-conversion (SPDC) [14, 15]. Notably, the demonstration of SPDC in two-dimensional layered materials offers significant advantages over bulk nonlinear materials both in terms of integration, due to their bond-free attachment to the underlying substrates [16], as well as in terms of relaxed phase-matching conditions, which need not be satisfied for subwavelength nonlinear active regions [17, 18].

Despite the significant interest in ultra-thin van der Waals materials for nonlinear applications, their small nonlinear conversion efficiency remains the strongest limiting factor. The reasons are the small interaction length inherent to significantly subwavelength nonlinear materials and, in many cases, the centrosymmetric crystalline structure, which constrains their second-order nonlinear susceptibility to be zero. This is particularly the case in TMDs and h-BN flakes formed by an even number of monolayers [4, 19, 20]. Furthermore, even for some noncentrosymmetric van der Waals materials, such as TMDs containing odd layers, a decrease in nonlinearity with increasing thickness is often observed, which is ascribed to the significant modification of the electronic structure driven by strong interlayer electronic coupling and dielectric screening [4].

Recently, efforts have been made to enhance the nonlinear effects in ultra-thin van der Waals materials either by breaking the crystal inversion symmetry through twisted flakes or by coupling them to photonic structures. In the case of the twisted flakes approach [21-23] one must resort to intricate flakes manipulations (e.g. the use of micro rotators), which is complex and renders the approach difficult to implement and incompatible with other basic functionalities. In the case of photonic resonators, two options can be considered. First, one can employ low cavity quality factor (Q) photonic structures, including metasurfaces and dielectric resonators, which enable a moderate enhancement but rather broadband operation [24-29]. Alternatively, one can resort to high Q resonant cavities that allow, potentially, for larger enhancements but sacrificing the broadband width due to their resonant character [30, 31]. This spectrally narrow enhancement prevents exploiting the full potentiality of subwavelength light sources, which can display a broad emission bandwidth as a result of the relaxed phase matching condition [14]. Thus, ideally one would like to merge the advantages of both approaches without necessitating complicated clean-room processes.

Herein we develop an approach to enhance nonlinear processes in ultra-thin materials via the engineering of the field distribution around the nonlinear material. This approach is independent of the material symmetry and, thus, of general applicability. We utilize nonlinear material/metal and nonlinear material/dielectric/metal heterostructure configurations to modify appropriately the electric field distribution at the van der Waals materials position. Our heterostructures play thus a double role: to enhance the intensity of the electric field while keeping a broadband response, similar to low Q cavities, and to enlarge the electric field gradient, which acts a second sizeable source of polarization. This second ingredient is essential to achieve a comparable nonlinear response between thin films with odd and even layers in materials with AA' stacking. To illustrate the general applicability of our heterostructures we chose two Van der Waals materials: hBN, which is a widespread 2D material, whose centrosymmetric or non-centrosymmetric character is layer number dependent, and $NbOCl_2$, which is a van der Waals material exhibiting one of the largest nonlinear responses to date (Supplementary Table 2). We observe a giant enhancement of second harmonic intensity on both h-BN, i.e. from some nanometres to tens of nanometre-thick layers, and $NbOCl_2$ flakes. Leveraging nonlinear transfer matrix simulations [23, 32] we have evaluated quantitatively the SH intensity on different heterostructures and substrates, getting insight into the interplay between dipolar and quadrupolar polarization contributions. Interestingly, for the SHG in h-BN layers the interplay between the two contributions results in a nonlinear intensity almost independent of the monolayers parity, erasing thereby one practical limitation. The nonlinear response from the developed structure is at least one order of magnitude (typically about 50 times) larger than on standard $SiO_2$/Si substrates and the observed enhancement effect is broadband, covering a wide range of pump laser wavelengths. This characteristic is particularly advantageous for ultra-thin van der Waals biphoton sources such as $NbOCl_2$, for which photon pairs exhibit a broad spectrum owing to relaxed phase-matching conditions [17, 33]. Finally, we propose and use an easy-to-implement $SiO_2$/Au planar structure that enables to enhance the nonlinear contributions in very thin (~20nm or less) nonlinear materials. With this method we achieve an impressive 3 orders of magnitude SHG enhancement in comparison with the conventional flakes on wafer configuration for h-BN layers with thicknesses below 10 nm. When employed to enhance photon pair generation in $NbOCl_2$, the metal/dielectric heterostructure enables us to reduce the $NbOCl_2$ thickness down to 16 nm. This constitutes the realization of SPDC with one of the thinnest nonlinear media among currently reported SPDC sources (Supplementary Table 4).

# Results

## General strategy for enhancing SHG in thin materials

The nonlinear optical processes in a material are governed by its electrical polarization, which is given by a power series of the electric field amplitude. Beyond the linear term [34]:

$P_{NLO} = P_{2\omega} + P_{3\omega} + \cdots = \chi^{(2)} E_\omega E_\omega + \chi^{(3)} E_\omega E_\omega E_\omega + \cdots$, where $\chi^{(2)}$ is the second-order susceptibility tensor. $\chi^{(2)}$ is at the basis of SHG, which not only holds potential for creating nonlinear devices but also is being employed extensively by the 2D community for non-destructive lattice orientation identification using low pump powers [4, 35, 36]. In the classical SHG process, two photons of frequency $\omega$ interact inside the nonlinear material to give a single photon of frequency $2\omega$. As indicated above, the SHG intensity is related to the pump electric field by the second-order susceptibility tensor that, in turn, can be expanded in terms of its dipole and higher multipole moments. To leading order, we can thus write $P_{2\omega} = \chi_d^{(2)} : E_\omega E_\omega + \chi_q^{(2)} : E_\omega \nabla E_\omega + \cdots$, where $\chi_d^{(2)}$ and $\chi_q^{(2)}$ represent the dipolar and quadrupolar moments of $\chi^{(2)}$[23, 37]. Note that the dipolar term couples only to the electric field amplitude (in fact, to the electric field intensity), while the quadrupolar term couples to the electric field amplitude and its gradient (i.e. the spatial variation of the electric field). For a more detailed analysis of this expansion and the quantitative relationship between the dipole and quadrupole moments see note "Dipole moment and Quadruple moment" in the supplementary information. Previous studies have demonstrated quadrupolar enhancement of SHG by using optical dressing with an in-plane photon wave vector (i.e. by breaking inversion symmetry by oblique incidence excitation) [38] and a two-beam SHG technique on a rough metal surface, on which unequal retardation effects for each beam breaks the initial symmetry [39].

Our strategy consists in finding simple designs that maximize the tradeoff between the electric field amplitude at the nonlinear-material location and the gradient of the electric field at that same position, as illustrated in Figure 1a. To do so we exploit two well-known facts of metals: first, they display a large reflectivity over a broad wavelength range in the visible and infrared (IR), comparable to Bragg mirrors with few number of pairs or with small refractive index contrast between the Bragg materials (i.e. those forming low Q Fabry-Perot cavities) [40]; second, by imposing a near-zero electric field at their surface, they enable to create a strong electric field gradient near the metal, far exceeding for example the field gradient established in thick h-BN or $NbOCl_2$ by residual below-bandgap absorption.

To illustrate quantitatively the enhancement of both magnitudes, in Figure 1b (left axis) we have plotted the electric field intensity inside 100 nm-thick h-BN deposited on gold and on standard $SiO_2$/Si wafers. Due to the relatively large reflectivity of the h-BN/gold interface compared to the h-BN/$SiO_2$ interface (Supplementary Table 5), the electric field intensity (at the pump wavelength) increases by a factor 3-5, providing a stronger dipolar contribution to the nonlinear polarization. Interestingly, the electric field gradient does also become larger when considering h-BN deposited on gold (right axis in Figure 1b), with an increase that can amount up to a factor ten compared to h-BN/$SiO_2$. This increased gradient, combined with the enhanced electric field amplitude, magnify the quadrupolar contribution to the nonlinear polarization.

Still, one can wonder how important the quadrupolar contribution can be, in absolute terms, compared to the dipolar one. In centrosymmetric materials, for which the dipolar contribution vanishes due to symmetry constraints, the quadrupolar contribution is obviously the dominant one, and it enables to observe SHG in sufficiently thick materials across which a field amplitude gradient can be established. On the other hand, in non-centrosymmetric materials the dipolar contribution is in general considered to be the dominant one. To illustrate the actual magnitude of the quadrupolar contribution enabled by the use of metals, we used the nonlinear transfer matrix approach [23] to calculate the individual contributions to the SHG intensity of h-BN displaying an odd number of monolayers (i.e. for a non-centrosymmetric BN) deposited on gold. Figure 1c indicates that when using gold as a substrate, the quadrupolar response in h-BN is just a factor 2 to 3 times smaller than the dipolar response. The fact that both contributions are of the same order of magnitude highlights the potential effect of the quadrupolar moment in modifying the nonlinear response of a given material, particularly of centrosymmetric materials, where the quadrupolar contribution dominates. Furthermore, because the global enhancement of dipolar and quadrupolar contributions described above stems from two general metal properties—high reflectivity over a broad wavelength range and the ability to generate a strong electric field gradient near the metal—its effect is observed when depositing the nonlinear active material on a variety of metals (Supplementary Figure 1, 16 and 17).

### Enhancing SHG of h-BN on metals and polarization properties

To illustrate the benefits of our simple heterostructure for SHG, we transferred h-BN flakes of different thicknesses onto $SiO_2$/Si substrates that were partially coated with 200 nm (Au)/6 nm (Ti) films (Supplementary Figure 3). Note that Ti is used here to facilitate metal adhesion and plays no particular role in terms of the optical design. A femtosecond-pulsed laser with tunable wavelength around 890 nm was used to conduct the nonlinear measurements on h-BN. More detailed description of the setup can be found in the Methods Section.

Figure 2(a) displays the SHG response of h-BN thin films calculated thanks to the nonlinear transfer matrix formalism (Section "Nonlinear transfer matrix method" in Supplementary Information) as a function of h-BN thickness, for odd (solid) and even (dashed) number of monolayers h-BN, on gold and $SiO_2$/Si substrates. The one to two orders of magnitude enhancement for even number of monolayers h-BN can be predominantly ascribed to the enhancement of both the amplitude of the electric field, due to an increased reflectivity at the h-BN/gold interface with respect to h-BN/$SiO_2$ interface, and its gradient, which both contribute to a larger quadrupolar response. On the other hand, for odd number of monolayers, the potential enhancement if only the dipolar contribution was considered would be typically ~~two~~ one to two orders of magnitude (attaining its maximum for an h-BN thickness of about 30 nm, for which the dipolar response on $SiO_2/Si$ approaches zero). However, due to the out-of-phase quadrupolar contribution (see Supplementary Figure 2), the overall enhancement is rather a factor 20-30.

Thus, to enhance the SHG response of even number of monolayers h-BN (i.e. centrosymmetric) by one to two orders of magnitude, we need to sacrifice a factor ~2-4 in the SHG signal of odd number of monolayers h-BN (i.e. non-centrosymmetric). Importantly, the enhancement of both contributions on gold substrates results in two SHG intensity curves (Figure 2a, red curves) that are almost parallel to each other and that differ at most by a factor of 2 for thicknesses larger than 15 nm, erasing thereby the strong dependence of monolayer parity observed on $SiO_2/Si$ wafers (Figure 2a, blue curves).

The experimental data in Figure 2a, while showing a large dispersion, confirm that employing gold as a substrate systematically enhances the SHG intensity of h-BN by approximately a factor of ~10. For thicknesses around 30 nm, this enhancement can exceed two orders of magnitude. More importantly, the SHG enhancement exhibits a more than 100 nm wide spectral range based on different excitations wavelengths, going from 395 nm to 445 nm, as illustrated in Figure 2b. This broad spectral range results from the broadband metal reflectivity at visible and IR wavelengths, and from the amplitude field being zero close to the metal surface, imposing thereby a strong field amplitude gradient for all considered wavelengths.

To further ensure that the SHG response arises from the h-BN we performed polarization-dependent SHG measurements, which exhibit a distinct six-fold symmetry (Figure 2c), in agreement with the $D_{3h}$ point group of h-BN and some TMDs [4, 21, 41]. To discard any potential contribution to the SHG signal from the metal surface nonlinearity, SHG was recorded under the same conditions from bare gold. The SHG response from the bare metal is about 20 and 80 times smaller than the SHG signal from h-BN/gold heterostructures with h-BN thicknesses of 36 nm and 65 nm, respectively, as shown by polarization-resolved measurements on these two h-BN flakes at different locations (Figures S4, S5 and S6). These findings corroborate that the measured SHG signal indeed originates mostly from the h-BN flakes. Furthermore, since the measurements on $SiO_2$/Si and gold substrates were performed on exactly the same h-BN flakes (Supplementary Figure 3), the azimuthal orientation of the maxima/minima in the polarization-resolved measurements for a given thickness coincide on both substrates (Figure 2c), as dictated by the common crystallographic orientation with respect to the polarized incident laser. Noticeably, the polar plot of the SHG is fully symmetric on the flat $SiO_2$/Si substrate while an asymmetry is observed systematically on gold films. This asymmetry degree varies, however, from location to location within the same flake (Figures S4 and S5). This observation is tentatively ascribed to the inhomogeneous strain induced by the irregularities of the gold film whose roughness, while in the nanometer scale, is five times larger than that of $SiO_2$/Si substrates (root mean square roughness of 2.4 nm Vs 0.5 nm, respectively). To minimize the effect of the polarization-resolved SHG response asymmetry of h-BN on gold, each experimental point in Figure 2a corresponds to the SHG value averaged over the six maxima in the polarization-resolved measurements, both for data on gold and on $SiO_2$/Si substrates.

Overall, the results in Figure 2 confirm that the SHG intensity of h-BN on gold originates from the h-BN and that it is enhanced by one to two orders of magnitude compared to h-BN on $SiO_2$/Si thanks to the interplay between enhanced dipolar and quadrupolar contributions.

## Compatibility of SHG metal and 2D twisting enhancements

To further tailor the nonlinear response and symmetry breaking of h-BN we combine our material-on-metal approach with twist engineering, which is known to break inversion symmetry by physically rotating h-BN sections with respect to each other. To do so we investigated the SHG of h-BN homostructures with two different stacking angles on gold and on $SiO_2$/Si substrates (Figures 3a, 3b, 3c, 3d, 3e and 3f). Specifically, we selected a 49 nm thick h-BN flake and separated it into two pieces. One piece of 49 nm h-BN layer was picked up while the other piece of 49 nm h-BN layer, left on the initial substrate, was rotated by 180° (equivalently, 60°) before stacking it side by side with the unrotated one. This pair of rotated h-BN layers were subsequently and simultaneously transferred on top of a third 46 nm h-BN layer, which had a rotation of 0° with respect to the two top initial layers (Figure 3a). In this way, h-BN homostructures with 0°- and 60°-rotation interfaces can be compared within the same sample and under exactly the same experimental conditions, as evidenced by the polarization-dependent SHG (Figure 3c).

The presence of an additional dipolar SHG signal arising from the h-BN interface at which the local inversion symmetry is broken by the layers twist [23] leads to a notable increase of SH intensity, magnified by approximately 357 times, compared to the off-gold AA' stack (Figures 3e and f). Note that the value of the enhancement depends on the location of the non-centrosymmetric interface within the entire structure (see numerical simulations in Figures S7 and S8), and can be engineered to accumulate in-phase the individual effects of subsequent AB interfaces distributed along the stack thickness [22, 23]. Besides, an optimum choice of the top and bottom h-BN pieces parity for the actual h-BN flakes would result in a further improvement (18 times more for 95 nm h-BN flakes, see Supplementary Figure 9b, 9d and 9f).

In addition to the enhanced bulk effects from dipolar and quadrupolar contributions -arising due to the increased electric field magnitude and field gradient-, the additional twisted interface-related SHG further amplifies the overall SHG response. This mechanism is certainly interesting and an additional source of SHG enhancement compatible with our approach.

However, it is also much more complicated, which highlights our approach of substrate engineering as a powerful and simple way of enhancing the nonlinear signals from van der Waals integrated materials.

## SPDC enhancement in nonlinear-material/metal structures

In ultrathin materials (largely subwavelength) phase-matching conditions can be assumed to be always fulfilled and, thus, one can expect SPDC to provide pairs of photons spreading over a broad wavelength range [14, 17, 18]. Furthermore, under certain conditions a correspondence between sum-frequency generation (SFG) and the SPDC intensity can be established [42], meaning that the larger the SFG intensity is, the larger the SPDC rate one can expect. Because of these two reasons, our material-on-metal approach based on broadband enhancement of SHG should be of use with subwavelength SPDC nonlinear materials.

To illustrate the potential of our approach we used $NbOCl_2$ as the nonlinear van der Waals material due to its high nonlinear coefficient (Supplementary Table 2) and potential for subwavelength SPDC emission [14]. We thus transferred an $NbOCl_2$ flake (275 nm thick) onto a gold film. A continuous-wave (CW) laser centered at 409 nm was used as the fundamental wave and a Hanbury Brown-Twiss setup [43] was used for all coincidence counting experiments. The SPDC response from $NbOCl_2$ was first measured when transferred onto an $SiO_2/Si$ substrate, showing a coincidence peak above the background when the time difference between the two detector channels is set to 0 ns, which is a signature of biphoton generation (Figure 4b). Interestingly, the $NbOCl_2$ flake on gold exhibits a ten-fold increase in the coincidence count rate. The polarization-independent characteristic of the Au film obviates the need for orientation-specific alignment between the $NbOCl_2$ flake and its substrate, thereby not influencing the efficiency of the SPDC emission. Same as for the SHG process in previous sections, the presence of the metal just below the $NbOCl_2$ redefines the boundary conditions of the fundamental wave, inducing an enhancement of the electric field and higher-order contributions in the nonlinear process, while increasing also the reflectivity at the signal and idler wavelengths.

To evidence that the photon pairs are generated by SPDC, we examined several critical aspects of the correlated photon pair generation process. Both the signal and idler photons are emitted with polarization along the crystallographic $b$-axis, demonstrating that $NbOCl_2$ exhibits mainly type-0 SPDC emission (Figure 4c). Figure 4d shows the coincidence counts rate from $NbOCl_2$ on the gold film across different detection wavelength ranges, with all measurements conducted under the co-polarized configuration. Wider detection bandwidth results in larger coincidence rate, consistent with broadband biphoton generation due to relaxed phase-matched conditions [17]. On the other hand, the narrower the detection bandwidth the lesser the accidental coincidences, increasing thereby the coincidence-to-accidental ratio (CAR) [14, 44], as shown in Supplementary Figure 10. The observed biphoton coincidence rate is linearly proportional to the fundamental wave pump power, as shown in Figure 4e, while the CAR is inversely proportional to it due to the increased background photons (Figure 4f), which is a typical indication of nonclassical photon pair emission [17]. Finally, we analyzed the coincidence count rate relative to the polarization of the fundamental wave. When rotating the polarization of the fundamental wave with respect to the $NbOCl_2$ flake and measuring SPDC in the co-polarized setting, the generated pattern exhibits a two-lobed symmetry (Figure 4g), consistent with the $C_2$ crystal group characteristic of $NbOCl_2$ [45] and with our SHG experiments on $NbOCl_2$ (Figure 5f). Compared with recent approaches to enhance SPDC in $LiNbO_3$ layers, which made use of resonant metasurfaces etched either onto the $LiNbO_3$ [46] or onto a dielectric film atop [47], our approach is much easier to implement and paves the way to more efficient studies of photon frequency entanglement across a broader range of operation wavelengths.

## Pushing nonlinear efficiency of nanometric sources

Up to now, for the SHG measurements with h-BN we have dealt with layers several tens of nanometers in thickness. However, many applications (e.g. quantum sensing and quantum communication) and certainly ultimate miniaturization require nonlinear sources below 20 nm or even 10 nm thicknesses [48]. Unfortunately, if such thin layers were deposited onto the gold substrates discussed previously, although the SHG signal would be larger than on standard $SiO_2$/Si substrates (Figure 2a), it would be substantially smaller than if the h-BN layer was just double in thickness. As shown in Figure 2a, when the layer thickness becomes 15 nm, the SH signal reduction is approximately fivefold in comparison with a layer thickness of 30 nm.

To strengthen the nonlinear response of films several monolayers to tens of monolayers thick, we have designed a structure consisting of a $SiO_2$/Au/Si wafer stack that incorporates a planar dielectric Fabry-Perot cavity between the material and the metal (Figure 5a). Through nonlinear transfer matrix calculations, we anticipate a significant enhancement originating from a magnified dipolar and quadrupolar contributions to the polarization at the h-BN location, compared to the signal obtained on the gold film directly, for h-BN flakes containing both odd and even numbers of monolayers (Figure 5b and Supplementary Figure 11). In the case of very thin layers, where the gradient of the electric field is smaller, the quantitative enhancement of SHG becomes again parity dependent. Very thin flakes with an odd number of monolayers tend to favor SHG over even-layered flakes, and optimized structures need to be fabricated depending on parity (Figure 5c). Note that for the current Air/$SiO_2$/Gold configuration, if we assume that the reflection phases at each of the two interfaces differ roughly by $\pi$, the minimum resonant cavity thickness is not $\lambda/2n_{cav}$ but $\lambda/4n_{cav}$. This resonant condition would correspond to a thickness in the order of 153 nm for the 890 nm pump and about 77 nm for the SH waves. For a monolayer-thick h-BN (i.e. for vanishingly-thin h-BN with odd number of monolayers), the optimum $SiO_2$ thicknesses are around ~76 nm and 166 nm (Figure 5b), so very close to the resonant condition for the SH waves and close to the condition for a maximum of the electric field pump amplitude (maximizing the dipolar contribution). Optimizing either of these two effects results in a similar computed SH signal. Furthermore, as shown in Figure 5b, for h-BN containing odd-number of monolayers and deposited on a $SiO_2$/gold substrate, the h-BN thickness maximizing the SH can be extremely thin (in the order of monolayers).

On the contrary, for h-BN with vanishingly-small even number of monolayers (Supplementary Figure 11), the optimum $SiO_2$ thickness (121 nm) lies between the resonant conditions for the SH waves and the pump. Besides, contrary to h-BN containing odd number of monolayers, very thin even-parity h-BN does not maximize the SHG response. Indeed, to maximize the quadrupolar moment controlling the response of centrosymmetric h-BN we need to have a nonnegligible h-BN thickness, leading to an optimum $SiO_2$ thickness of 51 nm for an optimum h-BN thickness of 59 nm.

To experimentally validate our findings, we transferred an 8 nm h-BN flake onto a 64 nm thick $SiO_2$ layer deposited on gold. The measured two-orders of magnitude enhancement of the SH signal compared to the same flake deposited directly onto gold (Figure 5d) confirms the efficiency of the $SiO_2$/Au structure in enhancing the nonlinear response of very thin h-BN, which still remains broadband (Supplementary Figure 12). Furthermore, this straightforward and versatile method can also be extended to enhance the SHG for thicker h-BN layers, providing three orders of magnitude enhancement compared to standard h-BN on $SiO_2$/Si substrate (red curve in Figure 5e). This is of practical importance for the 2D community to analyze the parity of thick h-BN samples in which spatial thickness variations in the order of the monolayer are expected, as illustrated in Supplementary Figure 13a. Monitoring of the SH photon count rate after the transfer of a monolayer flake onto a "thick" h-BN revealed a pronounced distinction between the SHG intensity of the 17 nm flake and its augmented state post-monolayer addition (Supplementary Figure 13b). This stark intensity contrast not only underscores the impact of the additional layer but also confirms the even-layered structure of the flake, consistent with our simulations. In comparison to the threefold contrast reported in previous studies for similar layer thickness differences [23], our $SiO_2$/Au structure approach markedly improves the contrast ratio by an order of magnitude.

To verify the applicability of the approach for enhancing quantum nonlinear processes too, we transferred a 16 nm $NbOCl_2$ flake onto the same 64 nm $SiO_2$/Au/Si wafer structure. The characterization of the classical SHG process, under co-polarized configuration measurement, displays a giant enhancement of over 300 times compared to 16 nm $NbOCl_2$ deposited on $SiO_2$/Si wafer (Figure 5f). The coincidences per second measured from the 16 nm $NbOCl_2$ flake on 64 nm $SiO_2$/Au/Si structure is presented in Figure 4g. A distinct coincidence count peak, with a CAR exceeding 8 (Supplementary Figure 14), was detected in the co-polarized configuration with a 0.25 mW pump power. A lower bound enhancement factor of 100 was obtained in a 100-minute measurement. To our knowledge, this structure features a nonlinear medium that is three times thinner than previously reported values [14], paving the way to the development of real ultra-thin integrated quantum devices. A quantitative comparison of "thin" active regions showing SPDC is provided in Supplementary Table 4.

## Discussion

We have proposed an easy-to-implement method for enhancing the nonlinear response of both centrosymmetric and non-centrosymmetric van der Waals materials that involves manipulating the amplitude, phase, as well as gradient of the pump field. By engineering the light component we achieved a substantial boost in the light-matter interaction resulting in a notable increase in the nonlinear response, as observed for both SHG and SPDC (Supplementary Table 6). For centrosymmetric materials like h-BN, which typically exhibit lower second-order susceptibility values due to their symmetry, our approach has enabled an impressive increase of about two orders of magnitude in the SHG response and enabled even-layer h-BN to exhibit SHG responses comparable to those of odd-layer h-BN. This demonstrates that the symmetry constraints on the nonlinear optical response can be effectively mitigated by adequately modifying the electric field distribution within the nonlinear material.

The desired operational bandwidth of a nonlinear optical system is highly dependent on the specific application. For instance, quantum memory of single photons requires an extremely narrow bandwidth to ensure high fidelity and efficiency in photon storage and retrieval. This contrasts with other applications, such as broadband optical parametric amplifiers and supercontinuum generation, which rely on robust operation across a wide range of frequencies and benefit from broader bandwidths. In addition, the growing interest in ultra-thin entangled photon sources, particularly for integrated quantum computing and communication devices, highlights the importance of broadband capabilities. Frequency entanglement, which leverages the relaxed momentum conservation in thin films, is increasingly recognized for its potential in encoding information. In such systems, it is essential to enhance the generation of photon pairs over a broad spectrum of frequencies to maximize entanglement efficiency. Our material-on-metal configurations are particularly well-suited for such broadband operations, with at least 100 nm large bandwidth. The broadband enhancement of quantum nonlinear processes was demonstrated by achieving a ten-fold increase in the SPDC coincidence rate of $NbOCl_2$ layers deposited on gold.

By introducing a $SiO_2$ layer between the van der Waals flakes and the gold film the enhancement for monolayer thick h-BN can be boosted by 3 orders of magnitude compared to $SiO_2$/Si. Thus, this method can accommodate h-BN layers ranging from the monolayer level to hundred nanometers thick. Most importantly, we demonstrate correlated photon pair generation in a 16 nm-thick $NbOCl_2$ layer transferred onto the same $SiO_2$/Au structure, suggesting its potential for achieving SPDC down to the monolayer level.

# Methods

The hexagonal boron nitride (h-BN) flakes were prepared by mechanical exfoliation with poly-dimethyl Siloxane (PDMS). The twisted bilayer h-BN was fabricated by the tear and stack method. The bottom layer h-BN was picked up by polycarbonate (PC) film and transferred onto the pre-patterned gold lead. Then another h-BN flake was torn apart partially and picked up by PC film at 70°C, leaving the other half of the h-BN on the $SiO_2$ substrate. The PC/h-BN stack on the transfer plate was rotated by 180° and stacked on top of the remaining half h-BN. Finally, the twisted h-BN bilayer was picked up at 70°C and dropped on top of the prepared h-BN flake on the gold to finish the whole structure. The PC film was removed with chloroform wash (Figure 3a).

The $NbOCl_2$ layers were obtained by mechanical exfoliation of bulk $NbOCl_2$. After that, the exfoliated flakes were transferred onto a polydimethylsiloxane (PDMS) film. We selected appropriate flakes on the PDMS film for the fabrication of the heterostructures.

As determined by AFM, the thickness of gold/titanium substrate used in Figures 2, 3 and 4 is 200 nm/18 nm, and the thickness of $SiO_2$ in the $SiO_2$/Au heterostructure of Figure 5 is 64 nm. The primary role of the Ti layer is to act as an adhesion promoter, ensuring a strong bond between the gold film and the underlying substrate. The thickness of $SiO_2$ layer on $SiO_2$/Si substrate is 285 nm. The thickness of $NbOCl_2$ shown in Figure 4 is 275 nm while in Figure 5 it is 16 nm.

Second harmonic generation

To perform second harmonic generation (SHG) measurements we utilized a custom-built confocal setup. The samples under investigation were excited using a tunable Ti: sapphire laser. The laser beam passed through a series of optical elements, including a polarizer, a half waveplate, a quarter waveplate, a beam splitter, a second half waveplate, and an objective. The emitted signals from the sample were directed through the beam splitter, a half waveplate, and a 600 nm shortpass filter before being directed to the spectrometer for analysis. Additionally, the half waveplate positioned next to the objective was motorized, allowing for investigations of angle-dependent SHG.

To ensure a more precise and reliable calculation of the SHG enhancement factor, we have utilized Lorentz fitting directly on the raw data to extract the maximum SHG signal values, ensuring that no artificial baseline adjustments impact the analysis. This method allows us to accurately capture the polarization-dependent SHG response for h-BN on different substrates, including gold and $SiO_2$. For each substrate, we identify six maximum values that correspond to the six-fold symmetry characteristic of h-BN. By performing the same measurements on h-BN on and off the gold substrates, we calculate the enhancement factor at each angle. We then compute the mean value and standard deviation to provide a robust estimate of the enhancement factor and its associated uncertainty. This approach ensures that the enhancement factor is derived from the true physical properties of the system, free from the influence of any background subtraction, and accurately reflects the influence of different substrates on the SHG response.

## Coincidence experiments

In our study, photon coincidence measurements were conducted using a specialized Hanbury Brown-Twiss setup. A 409 nm continuous wave (CW) laser was used to excite our samples, utilizing a back-reflection configuration. In the excitation path, the laser passed through a polarizer, a half waveplate, a quarter waveplate, a dichroic mirror, another motorized half waveplate, and a microscope objective (50x magnification, 0.8 numerical aperture). The emitted signals from the sample passed through the same objective, dichroic mirror, a 750 nm long-pass filter, a beam splitter and then were divided into two output channels. Each channel was equipped with a quarter waveplate, a half waveplate and a polarizer before being directed to the avalanche photodiode for analysis. The correlation signals from the two paths are identified by a time tagger (Swabian Instruments). We set the polarization of the fundamental waves aligned to the orientation of the crystallographic polar axis of the $NbOCl_2$ (denoted as b). The lower bound of the enhancement factor for SPDC is determined using the signal-to-noise ratio (SNR), as the SPDC signal from the material off the gold film is at or below the background level. Specifically, it is calculated as: $(C^{on} - \langle A \rangle^{on})/\delta A^{on}$ and $(C^{on} - \langle A \rangle^{on})/\delta A^{off}$, where $C$ is the coincidence counts per second and $A$ is the accidental counts per second for on and off structures, $\langle \rangle$ means mean-value and $\delta$ represents the standard deviation. Using this approach, the calculated lower bound values of the enhancement factor are 93 and 153, respectively, confirming a significant enhancement of the SPDC signal. The bin width in our coincidence experiments was set to 5000 ps.

## Nonlinear transfer matrix method calculations

To understand SHG enhancement we use nonlinear transfer matrix simulations [23, 32] in which the only approximation is that the pump field is unaffected by the second order nonlinear process, the so-called undepleted-pump approximation. This formalism enables us to calculate precisely the pump field distribution (i.e. its spatially varying amplitude as well as its spatially-varying gradient, as illustrated in Figure 1a) imposed by the different interfaces and materials employed, especially on metal surfaces. It also allows us to introduce the nonlinear polarization sources at the level of each h-BN monolayer, facilitating the analysis of the effect of the actual parity of monolayers (odd or even) as well as of rotated heterostructures [23, 49]. Noticeably, phase shifts between different waves involved in the process (pump, nonlinear sources and second-harmonic) are inherently included in this formalism. Thus, phase-matching conditions limiting SHG in bulk crystals are naturally retrieved.

Nonlinear Susceptibility Tensor of NbOCl2

The crystal structure of NbOCl$_2$ belongs to the C$_2$ space group, which determines the form of its second-order nonlinear susceptibility tensor:

$$\begin{bmatrix} 0 & 0 & 0 & d_{14}^{(2)} & 0 & d_{16}^{(2)} \\ d_{21}^{(2)} & d_{22}^{(2)} & d_{23}^{(2)} & 0 & d_{25}^{(2)} & 0 \\ 0 & 0 & 0 & d_{34}^{(2)} & 0 & d_{36}^{(2)} \end{bmatrix} \quad (1)$$

,where $d_{14}^{(2)} = d_{25}^{(2)} = d_{36}^{(2)}$, $d_{23}^{(2)} = d_{34}^{(2)}$, and $d_{16}^{(2)} = d_{21}^{(2)}$.

When the pump field is incident perpendicularly to the NbOCl$_2$ crystal (i.e. vertically in Figure 4a), the non-zero term contributing to the polarization in the nonlinear material is $d_{22}^{(2)}$ and $d_{23}^{(2)}$. Considering that $d_{22}^{(2)} >> d_{23}^{(2)}$, the main SPDC response in NbOCl$_2$ is type-0. In type-0 SPDC, the polarization of the pump, signal, and idler photons are all the same.

# Data availability

Relevant data supporting the key findings of this study are available within the article and the Supplementary Information file. All raw data generated during the current study are available from the corresponding authors upon request.

# Additional information

Correspondence and requests for material should be addressed to Jesus Perez (jesus.zuniga@ntu.edu.sg) and Wei-bo Gao (wbgao@ntu.edu.sg).

# References

1. Liu, M., et al., *A graphene-based broadband optical modulator.* Nature, 2011. **474**(7349): p. 64-67.
2. Gan, X., et al., *Chip-integrated ultrafast graphene photodetector with high responsivity.* Nature photonics, 2013. **7**(11): p. 883-887.
3. Koppens, F., et al., *Photodetectors based on graphene, other two-dimensional materials and hybrid systems.* Nature nanotechnology, 2014. **9**(10): p. 780-793.
4. Li, Y., et al., *Probing symmetry properties of few-layer MoS2 and h-BN by optical second-harmonic generation.* Nano letters, 2013. **13**(7): p. 3329-3333.

5.      Malard, L.M., et al., *Observation of intense second harmonic generation from MoS 2 atomic crystals.* Physical Review B, 2013. **87**(20): p. 201401.

6.      Weismann, M. and N.C. Panoiu, *Theoretical and computational analysis of second-and third-harmonic generation in periodically patterned graphene and transition-metal dichalcogenide monolayers.* Physical Review B, 2016. **94**(3): p. 035435.

7.      Klein, J., et al., *Electric-field switchable second-harmonic generation in bilayer MoS2 by inversion symmetry breaking.* Nano letters, 2017. **17**(1): p. 392-398.

8.      Paradisanos, I., et al., *Second harmonic generation control in twisted bilayers of transition metal dichalcogenides.* Physical Review B, 2022. **105**(11): p. 115420.

9.      Autere, A., et al., *Nonlinear optics with 2D layered materials.* Advanced Materials, 2018. **30**(24): p. 1705963.

10.     Srivastava, Y.K., et al., *MoS2 for Ultrafast All-Optical Switching and Modulation of THz Fano Metaphotonic Devices.* Advanced Optical Materials, 2017. **5**(23): p. 1700762.

11.     Wu, K., et al., *All-optical phase shifter and switch near 1550nm using tungsten disulfide (WS2) deposited tapered fiber.* Optics Express, 2017. **25**(15): p. 17639-17649.

12.     Trovatello, C., et al., *Optical parametric amplification by monolayer transition metal dichalcogenides.* Nature Photonics, 2021. **15**(1): p. 6-10.

13.     Ciattoni, A., et al., *Phase-matching-free parametric oscillators based on two-dimensional semiconductors.* Light: Science & Applications, 2018. **7**(1): p. 5.

14.     Guo, Q., et al., *Ultrathin quantum light source with van der Waals NbOCl2 crystal.* Nature, 2023. **613**(7942): p. 53-59.

15.     Weissflog, M.A., et al., *A tunable transition metal dichalcogenide entangled photon-pair source.* Nature Communications, 2024. **15**(1): p. 7600.

16.     Liu, Y., Y. Huang, and X. Duan, *Van der Waals integration before and beyond two-dimensional materials.* Nature, 2019. **567**(7748): p. 323-333.

17.     Santiago-Cruz, T., et al., *Entangled photons from subwavelength nonlinear films.* Optics Letters, 2021. **46**(3): p. 653-656.

18.     Okoth, C., et al., *Microscale Generation of Entangled Photons without Momentum Conservation.* Physical Review Letters, 2019. **123**(26): p. 263602.

19.     Kim, S., et al., *Second-harmonic generation in multilayer hexagonal boron nitride flakes.* Optics letters, 2019. **44**(23): p. 5792-5795.

20.     Fryett, T., A. Zhan, and A. Majumdar, *Cavity nonlinear optics with layered materials.* Nanophotonics, 2017. **7**(2): p. 355-370.

21.     Kim, C.-J., et al., *Stacking order dependent second harmonic generation and topological defects in h-BN bilayers.* Nano letters, 2013. **13**(11): p. 5660-5665.

22.     Hong, H., et al., *Twist Phase Matching in Two-Dimensional Materials.* Physical Review Letters, 2023. **131**(23): p. 233801.

23.     Yao, K., et al., *Enhanced tunable second harmonic generation from twistable interfaces and vertical superlattices in boron nitride homostructures.* Science Advances, 2021. **7**(10): p. eabe8691.

24.     Gupta, T.D., et al., *Second harmonic generation in glass-based metasurfaces using tailored surface lattice resonances.* Nanophotonics, 2021. **10**(13): p. 3465-3475.

25.     Qu, L., et al., *Giant Second Harmonic Generation from Membrane Metasurfaces.* Nano Letters, 2022. **22**(23): p. 9652-9657.

26.     Kühner, L., et al., *High-Q nanophotonics over the full visible spectrum enabled by hexagonal boron nitride metasurfaces.* Advanced Materials, 2023. **35**(13): p. 2209688.

27.     Nauman, M., et al., *Tunable unidirectional nonlinear emission from transition-metal-dichalcogenide metasurfaces.* Nature Communications, 2021. **12**(1): p. 5597.

28. Timpu, F., et al., *Enhanced Nonlinear Yield from Barium Titanate Metasurface Down to the Near Ultraviolet.* Advanced Optical Materials, 2019. **7**(22): p. 1900936.

29. Popkova, A.A., et al., *Nonlinear Exciton-Mie Coupling in Transition Metal Dichalcogenide Nanoresonators.* Laser & Photonics Reviews, 2022. **16**(6): p. 2100604.

30. Bernhardt, N., et al., *Quasi-BIC Resonant Enhancement of Second-Harmonic Generation in WS2 Monolayers.* Nano Letters, 2020. **20**(7): p. 5309-5314.

31. Day, J.K., et al., *Microcavity enhanced second harmonic generation in 2D MoS2.* Optical Materials Express, 2016. **6**(7): p. 2360-2365.

32. Bethune, D., *Optical harmonic generation and mixing in multilayer media: analysis using optical transfer matrix techniques.* JOSA B, 1989. **6**(5): p. 910-916.

33. Sultanov, V., T. Santiago-Cruz, and M.V. Chekhova, *Flat-optics generation of broadband photon pairs with tunable polarization entanglement.* Optics Letters, 2022. **47**(15): p. 3872-3875.

34. Boyd, R.W., A.L. Gaeta, and E. Giese, *Nonlinear optics,* in *Springer Handbook of Atomic, Molecular, and Optical Physics.* 2008, Springer. p. 1097-1110.

35. Stepanov, E., et al., *Direct observation of incommensurate–commensurate transition in graphene-hBN heterostructures via optical second harmonic generation.* ACS applied materials & interfaces, 2020. **12**(24): p. 27758-27764.

36. Zhang, M., et al., *Emergent second-harmonic generation in van der Waals heterostructure of bilayer MoS2 and monolayer graphene.* Science Advances, 2023. **9**(11): p. eadf4571.

37. Torre, A.d.l., et al., *Mirror symmetry breaking in a model insulating cuprate.* Nature Physics, 2021. **17**(7): p. 777-781.

38. Zhang, Y., et al., *Doping-Induced Second-Harmonic Generation in Centrosymmetric Graphene from Quadrupole Response.* Physical Review Letters, 2019. **122**(4): p. 047401.

39. Wang, F.X., et al., *Enhancement of bulk-type multipolar second-harmonic generation arising from surface morphology of metals.* New Journal of Physics, 2010. **12**(6): p. 063009.

40. Médard, F., et al., *Experimental observation of strong light-matter coupling in ZnO microcavities: Influence of large excitonic absorption.* Physical Review B, 2009. **79**(12): p. 125302.

41. Mennel, L., M. Paur, and T. Mueller, *Second harmonic generation in strained transition metal dichalcogenide monolayers: MoS2, MoSe2, WS2, and WSe2.* APL Photonics, 2019. **4**(3).

42. Marino, G., et al., *Spontaneous photon-pair generation from a dielectric nanoantenna.* Optica, 2019. **6**(11): p. 1416-1422.

43. Brown, R.H. and R.Q. Twiss, *Correlation between photons in two coherent beams of light.* Nature, 1956. **177**(4497): p. 27-29.

44. Ivanova, O.A., et al., *Multiphoton correlations in parametric down-conversion and their measurement in the pulsed regime.* Quantum Electronics, 2006. **36**(10): p. 951.

45. Abdelwahab, I., et al., *Highly Efficient Sum-Frequency Generation in Niobium Oxydichloride NbOCl2 Nanosheets.* Advanced Optical Materials, 2023. **11**(7): p. 2202833.

46. Santiago-Cruz, T., et al., *Photon Pairs from Resonant Metasurfaces.* Nano Letters, 2021. **21**(10): p. 4423-4429.

47. Zhang, J., et al., *Spatially entangled photon pairs from lithium niobate nonlocal metasurfaces.* Science Advances, 2022. **8**(30): p. eabq4240.

48. Durand, A., et al., *Optically Active Spin Defects in Few-Layer Thick Hexagonal Boron Nitride.* Physical Review Letters, 2023. **131**(11): p. 116902.

49.   Hong, H., et al., *Giant enhancement of optical nonlinearity in two-dimensional materials by multiphoton-excitation resonance energy transfer from quantum dots.* Nature Photonics, 2021. **15**(7): p. 510-515.

## Acknowledgments

This work was supported by Singapore National Research foundation through CRP grants (CRP Award Nos. NRF-CRP22-2019-0004) and QEP Grants (NRF2021-QEP2-01-P01, NRF2021-QEP2-01-P02, NRF2021-QEP2-03-P01, NRF2022-QEP2-02-P14), and ASTAR IRG (M21K2c0116).

## Author Contributions Statement

Conceptualization, X.L., J.Z.P., W.G. Methodology, X.L., L.K., J.Z.P., W.G. Material, H.C., L.A., R.D., C.Z., Z.L. Simulation, X.L. Visualization, X.L. Discussion, S.J.W., Q.T., A.L. Writing – original draft, X.L., J.Z.P., W.G. Writing – review & editing, X.L., L.K., R.H., Y.M., J.Z.P., W.G.

## Competing Interests Statement

The authors declare no competing interests.

# Figure Legends/Captions

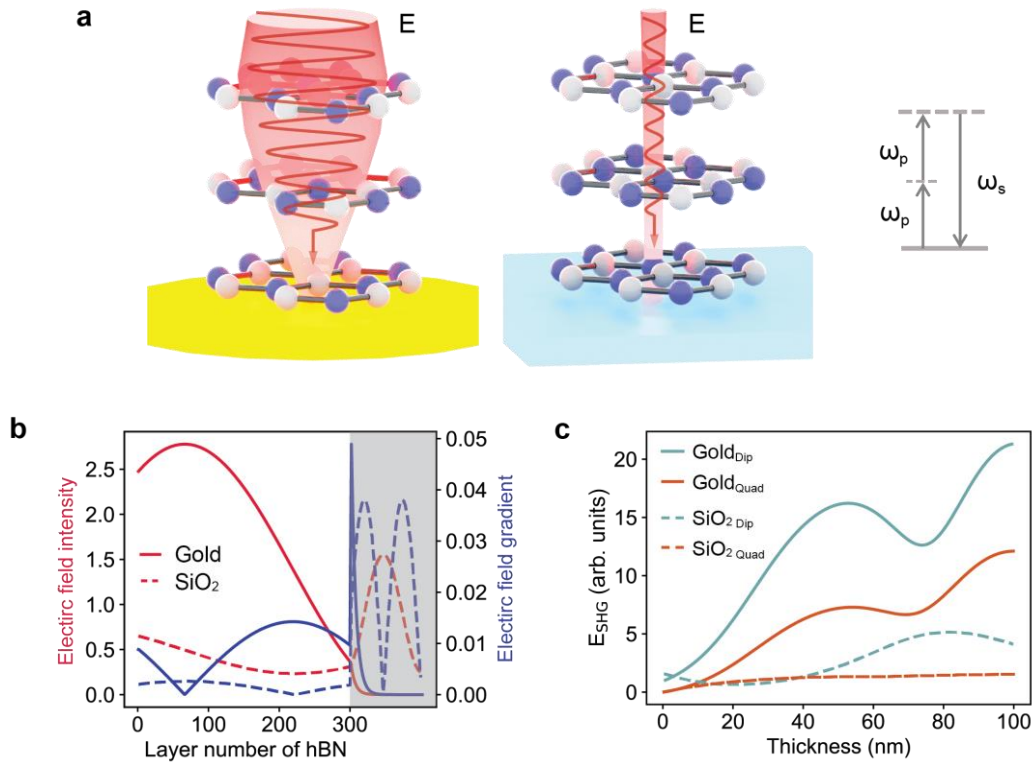

**Figure 1 | Electric field distribution in centrosymmetric-nonlinear material/metal heterostructures. a,** Schematic of incident light distribution on gold film (left) and SiO₂/Si substrate (right), showing a magnified electric field amplitude and gradient on the gold film. Inset: Illustration of the second harmonic generation (SHG) process, where two photons with frequency $\omega_p$ are combined to generate a single signal photon with frequency $\omega_s$. **b,** Simulation results of the magnitude of pump field intensity (red) and the gradient of electric field (blue) in a 300-monolayer hexagonal boron nitride (h-BN) flake on gold (line) and SiO₂/Si (dashed line). The dark area is the substrate. **c,** The simulation results of dipolar (green) and quadrupolar (orange) contributions to the magnitude of SHG electric field of h-BN flakes with odd number of monolayers on gold (line) as well as SiO₂/Si (dashed line). Dip: dipolar; Quad: quadrupolar.

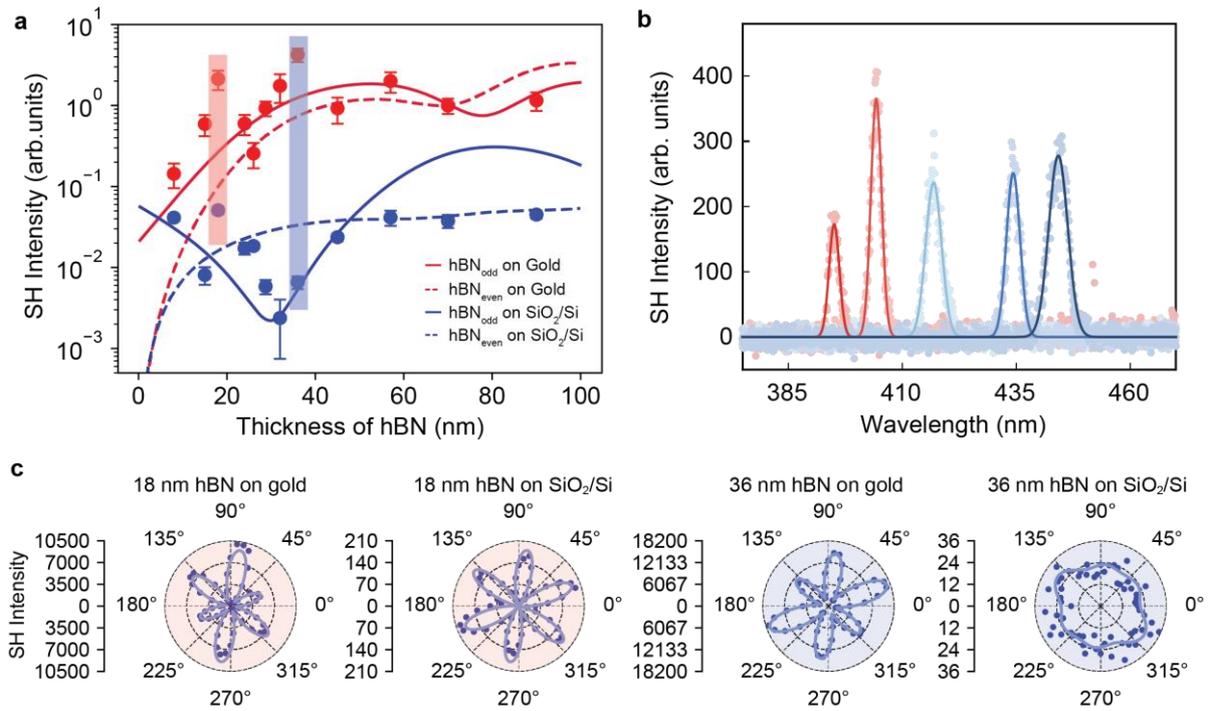

**Figure 2 | SHG enhancement in centrosymmetric-nonlinear material/metal heterostructures. a,** The experimental and simulated SH intensity for h-BN with different thicknesses on gold films and SiO$_2$/Si substrates. Symbols are experimental points while solid and dashed lines are nonlinear transfer matrix simulation results. Error bars represent the standard deviation of SHG values over the six maxima in the polarization-resolved measurements. The pump field in the simulation is 890 nm. All data are normalized to the value for 70 nm h-BN on gold. **b,** SHG of h-BN on gold film under variable pumping wavelengths at 790 nm, 808 nm, 834 nm, 868 nm, 890 nm. Solid lines represent Lorentzian fits to the SH signal spectrum. **c,** Polar plot of h-BN SH intensities as a function of the polarization angle of the incident laser on the same 18 nm (left) and 36 nm thick h-BN flakes (right) transferred onto a gold film and onto a SiO$_2$/Si substrate, respectively. The corresponding experimental points are highlighted by shadowed regions in Fig. 2a, where we assume that both flakes have an odd-layer structure. Solid lines represent fits to the polarization-dependent SHG response under strain.

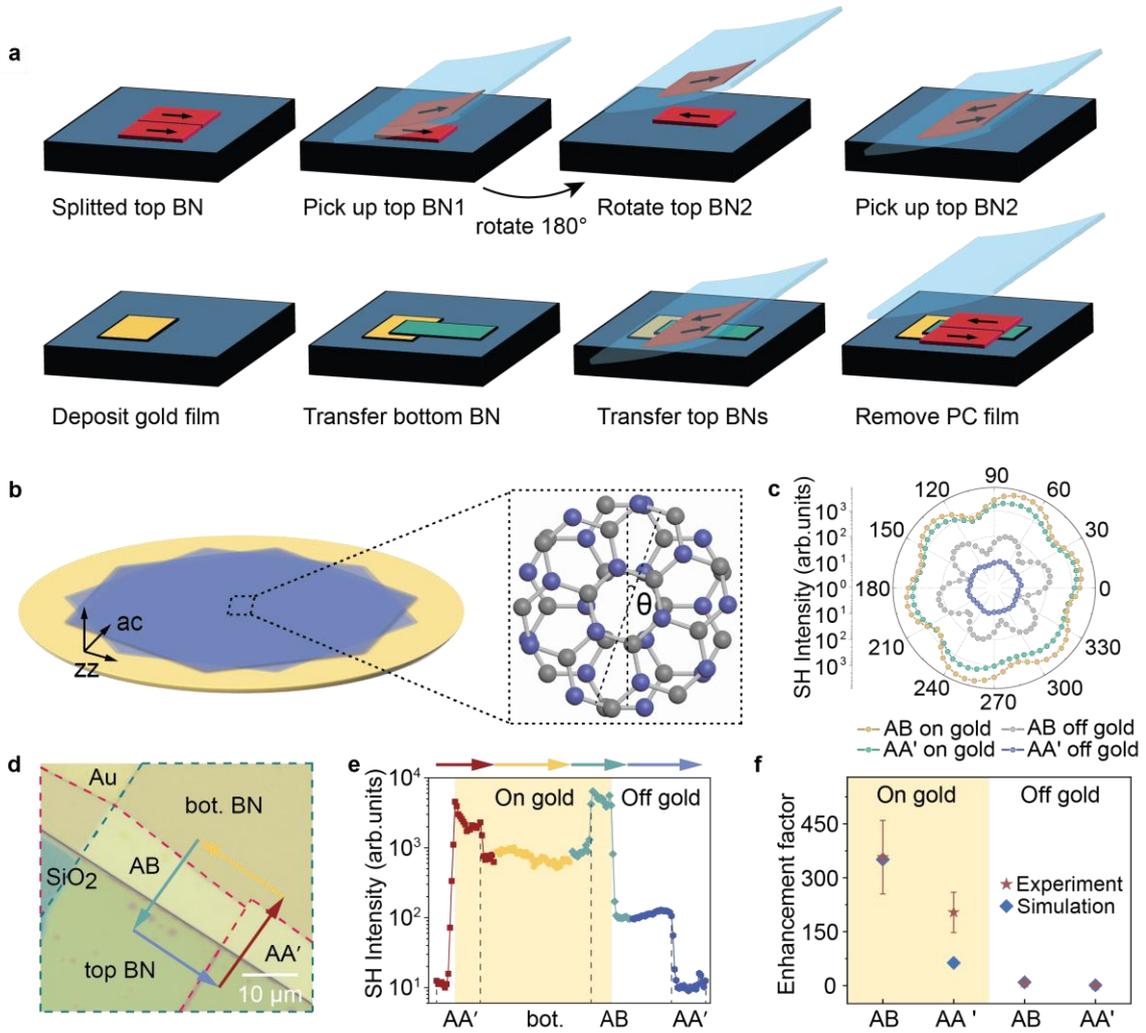

**Figure 3 | SHG enhancement in twisted h-BN on gold film. a,** Detailed process for preparing the h-BN homostructure. It illustrates the step-by-step method used to assemble the h-BN homostructure. PC: polycarbonate. **b,** Schematic of h-BN homostructures (blue) on gold film (yellow). A homostructure is formed by rotating the top h-BN flake relative to the bottom one by an angle. Inset: Rotation of h-BN lattice with a θ angle. **c,** The experimental results of SH intensity as a function of polarization angle for the homostructures on gold and SiO₂/Si; AA' stack corresponds to 0∘ and AB stack corresponds to 60∘ stacking angle. **d,** Bright field optical microscopy image of h-BN homostructures. Red dashed line: top layer h-BN. Green dashed line: bottom layer h-BN. **e,** SH intensity along a closed loop including all the h-BN homostructures (AA' and AB)/substrates(gold and SiO₂/Si) combinations. Different combinations are represented by unique colors and symbols, each corresponding to the path depicted by a line of the same color in **d**. The dashed lines mark the boundaries between different regions. **f,** Summary of the enhancement factors from different areas in **d** with respect to the 0° homostructure on SiO₂/Si substrates, measured at a 72° polarization angle: AB (60°) homostructure on gold; AA' (0°) homostructure on gold;

AB (60°) homostructure on SiO$_2$/Si; AA' (0°) homostructure on SiO$_2$/Si. The error bars represent the standard deviation of the enhancement factor within each region.

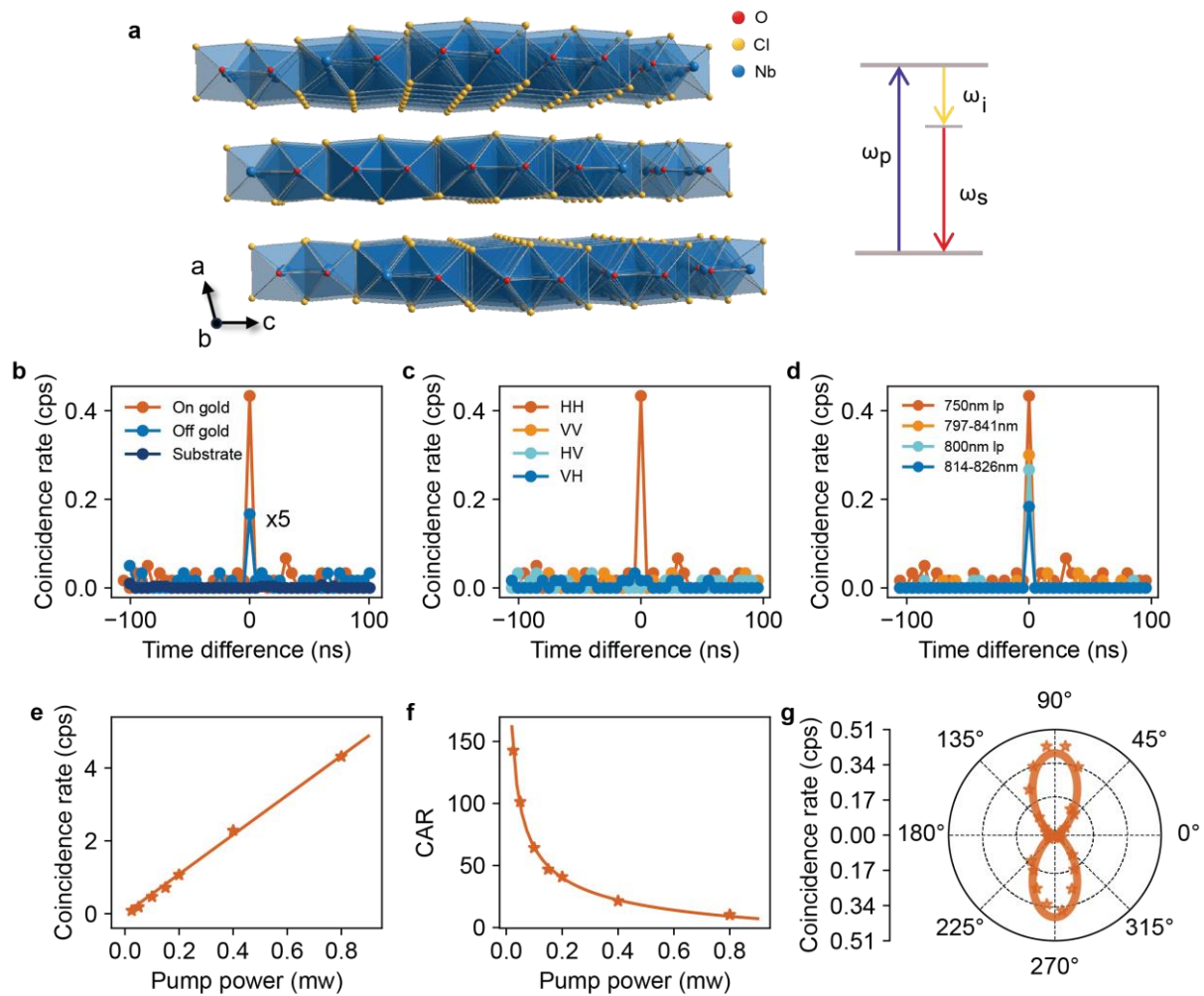

**Figure 4 | Characterization of photon pair generation from NbOCl$_2$/metal heterostructures. a,** Schematic drawing of the crystal structure of NbOCl$_2$, denoting the polar axis b, nonpolar axis c, and out-of-plane axis a. Red, oxygen atoms; yellow, chlorine atoms; blue, niobium atoms., Inset: An illustration of the quantum spontaneous parametric down conversion (SPDC) process, in which a photon ($\omega_p$) incident upon a nonlinear crystal is spontaneously converted into signal and idler photons of lower frequency ($\omega_i$ and $\omega_s$). **b,** Coincidence counts rate from 275nm thick NbOCl$_2$ on gold (orange), NbOCl$_2$ on SiO$_2$/Si (blue) and the bare SiO$_2$/Si substrate (dark blue). **c,** Coincidence rate in H-V basis from NbOCl$_2$ on gold substrate. H-axis is defined as the polar axis (b-axis), and the fundamental wave is also polarized along the H-axis. **d,** Coincidence counts rate from NbOCl$_2$ on gold film with different wavelength integration ranges. **e,** Power dependent coincidence counts rate from NbOCl$_2$ on gold film, showing a linear response. The solid line represents the fitted curve. **f,** Power dependent

coincidence-to-accidental ratio (CAR) from NbOCl$_2$ on gold film, indicating that a clear correlation peak above the classical limit (CAR > 2) is obtained from the sample. The solid line represents the fitted curve, which exhibits an inverse relationship. The results in **d-f** are measured under co-polarized configuration. **g,** Coincidence counts rate as a function of the fundamental wave polarization in the co-polarized detection configuration. Points are experimental results, and the broad line is a theoretical fitting.

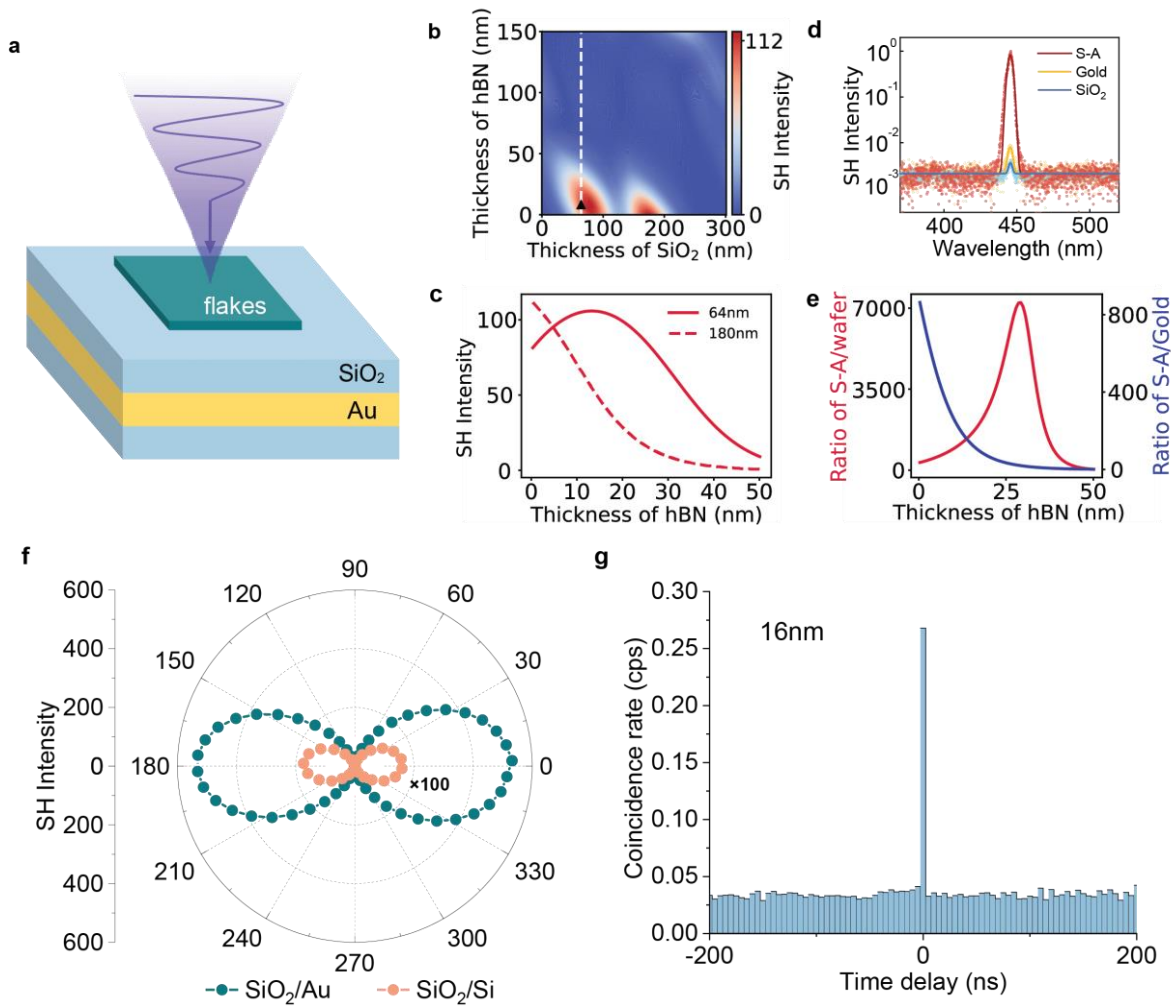

**Figure 5 | Characterization of enhanced nonlinear processes from few-monolayer flakes on dielectric/gold heterostructures. a,** The schematic of NbOCl$_2$ flake on SiO$_2$/Au structure. **b,** SH intensities as a function of h-BN and SiO$_2$ (on top of gold) thicknesses for h-BN flakes with an odd number of layers, deposited on a substrate made of a SiO$_2$/Au/Si wafer structure. The dashed line corresponds to the calculation range for the solid line shown in **c. c,** Simulated SH intensities for h-BN on the SiO2/Au/Si wafer structure, considering different h-BN and SiO$_2$ thicknesses: 64 nm for odd-layered h-BN (solid), 180 nm for odd-layered h-BN (dashed). **d,** Experimental results highlighting the SH intensities for an 8 nm h-BN flake on SiO$_2$-Au (red), gold (yellow),

SiO$_2$/Si (blue) substrate. **e**, Enhancement factor of SH intensity from h-BN flakes on 64 nm SiO$_2$-Au compared to Si wafer ($\frac{I_{S-A}^{SH}}{I_{wafer}^{SH}}$, red) and enhancement factor of SH intensity from h-BN flakes on 64 nm SiO$_2$-Au compared to gold ($\frac{I_{S-A}^{SH}}{I_{gold}^{SH}}$, blue), both involving h-BN flakes with an odd number of layers. **f**, Polarization dependent SHG measurement of 16 nm NbOCl$_2$ flake on 64 nm SiO$_2$/Au substrate (green) and SiO$_2$/Si wafer (orange). The 16nm NbOCl$_2$ flake on SiO$_2$/Au substrate shows an enhancement of SHG intensity by a factor 306. **g**, SPDC coincidence counts rate on 16 nm NbOCl$_2$ flake on 64 nm SiO$_2$/Au under co-polarized configuration with an enhancement factor 100 (lower bound) in a 100-minute measurement.